\documentclass[aps,twocolumn,superscriptaddress]
{revtex4-1}
\usepackage{amsfonts}
\usepackage{dcolumn}
\usepackage{bm}
\usepackage{tikz}
\usepackage{ulem}
\usepackage{filecontents}
\usepackage[colorlinks=true,citecolor=blue,urlcolor=blue]{hyperref}
\usepackage{amsmath,amsfonts,amssymb,times}
\usepackage{natbib} 
\setcitestyle{super,comma}

\makeatletter
\newcommand*{\Rom}[1]{\expandafter\@slowromancap \romannumeral #1@}
\makeatother

\begin{document}
\title{Observation of Non-Hermitian Spectral Deformation in Complex Momentum Space}

\author{Mu Yang}\thanks{These authors contribute equally to this work\label{Contribute}}
\affiliation{Laboratory of Quantum Information, University of Science and Technology of China, Hefei 230026, China}
\affiliation{Anhui Province Key Laboratory of Quantum Network, University of Science and Technology of China, Hefei, Anhui 230026, China}
\affiliation{CAS Center For Excellence in Quantum Information and Quantum Physics, University of Science and Technology of China, Hefei 230026, China}
\affiliation{Hefei National Laboratory, University of Science and Technology of China, Hefei 230088, China}

\author{Yue Li}\thanks{These authors contribute equally to this work\label{Contribute}}
\affiliation{Laboratory of Quantum Information, University of Science and Technology of China, Hefei 230026, China}
\affiliation{Anhui Province Key Laboratory of Quantum Network, University of Science and Technology of China, Hefei, Anhui 230026, China}
\affiliation{CAS Center For Excellence in Quantum Information and Quantum Physics, University of Science and Technology of China, Hefei 230026, China}
\affiliation{Hefei National Laboratory, University of Science and Technology of China, Hefei 230088, China}

\author{Mingtao Xu}
\affiliation{Laboratory of Quantum Information, University of Science and Technology of China, Hefei 230026, China}
\affiliation{Anhui Province Key Laboratory of Quantum Network, University of Science and Technology of China, Hefei, Anhui 230026, China}
\affiliation{CAS Center For Excellence in Quantum Information and Quantum Physics, University of Science and Technology of China, Hefei 230026, China}
\affiliation{Hefei National Laboratory, University of Science and Technology of China, Hefei 230088, China}

\author{Wei Yi}\email{wyiz@ustc.edu.cn}
\affiliation{Laboratory of Quantum Information, University of Science and Technology of China, Hefei 230026, China}
\affiliation{Anhui Province Key Laboratory of Quantum Network, University of Science and Technology of China, Hefei, Anhui 230026, China}
\affiliation{CAS Center For Excellence in Quantum Information and Quantum Physics, University of Science and Technology of China, Hefei 230026, China}
\affiliation{Hefei National Laboratory, University of Science and Technology of China, Hefei 230088, China}

\author{Jin-Shi Xu}\email{jsxu@ustc.edu.cn}
\affiliation{Laboratory of Quantum Information, University of Science and Technology of China, Hefei 230026, China}
\affiliation{Anhui Province Key Laboratory of Quantum Network, University of Science and Technology of China, Hefei, Anhui 230026, China}
\affiliation{CAS Center For Excellence in Quantum Information and Quantum Physics, University of Science and Technology of China, Hefei 230026, China}
\affiliation{Hefei National Laboratory, University of Science and Technology of China, Hefei 230088, China}

\author{Chuan-Feng Li}\email{cfli@ustc.edu.cn}
\affiliation{Laboratory of Quantum Information, University of Science and Technology of China, Hefei 230026, China}
\affiliation{Anhui Province Key Laboratory of Quantum Network, University of Science and Technology of China, Hefei, Anhui 230026, China}
\affiliation{CAS Center For Excellence in Quantum Information and Quantum Physics, University of Science and Technology of China, Hefei 230026, China}
\affiliation{Hefei National Laboratory, University of Science and Technology of China, Hefei 230088, China}

\author{Guang-Can Guo}
\affiliation{Laboratory of Quantum Information, University of Science and Technology of China, Hefei 230026, China}
\affiliation{Anhui Province Key Laboratory of Quantum Network, University of Science and Technology of China, Hefei, Anhui 230026, China}
\affiliation{CAS Center For Excellence in Quantum Information and Quantum Physics, University of Science and Technology of China, Hefei 230026, China}
\affiliation{Hefei National Laboratory, University of Science and Technology of China, Hefei 230088, China}

\begin{abstract}
Open systems feature a variety of phenomena that arise from non-Hermitian physics. 
Recent theoretical studies have offered much insights into these phenomena through the non-Bloch band theory, though many of the theory's key features are experimentally elusive. 
For instance, the correspondence between complex momenta and non-Hermitian bands, while central to non-Bloch band theory,
has so far defied direct experimental observation.
Here we experimentally study the non-Hermitian spectral deformation in complex-momentum space, by implementing a non-Hermitian lattice with long-range couplings in the synthetic orbital-angular-momentum (OAM) dimension of photons inside a degenerate cavity. 
Encoding the complex momenta in the phase and amplitude modulations of the OAM modes, and devising a complex-momentum-resolved projective detection, we reconstruct the spectral deformation in momentum space, where the eigenspectrum on the complex plane morphs through distinct geometries. 
This enables us to experimentally extract key information of the system 
under the non-Bloch band theory, including exceptional points in the complex-momentum space, the open-boundary spectra, and the generalized Brillouin zone.
Our work demonstrates a versatile platform for exploring non-Hermitian physics and non-Bloch band theory, and opens the avenue for direct experimental investigation of non-Bloch features in the complex-momentum space.
\end{abstract}

\maketitle

\section{Introduction}
Topological band theory~\cite{bansil2016colloquium} has profoundly reshaped our understanding of materials and provides geometric insights into topological matter such as Chern insulators~\cite{haldane1988model, liu2016quantum,chang2013experimental} and Weyl semimetals~\cite{armitage2018weyl,xu2015discovery}.
In recent years, growing attention has been directed toward band topology in non-Hermitian settings~\cite{shen2018topological,kawabata2019symmetry, gong2018topological}, where gain, loss, or nonreciprocal couplings render an effective non-Hermitian description for open systems.
In non-Hermitian systems, the geometry of the eigenspectrum can deform significantly on the complex plane under different boundary conditions, a phenomenon underlying the non-Hermitian skin effect~\cite{yao2018edge,okuma2020topological,kunst2018biorthogonal} and necessitating the application of non-Bloch band theory to restore the bulk-boundary correspondence~\cite{yao2018edge,yokomizo2019non,yao2018non,xiao2020non,zhang2020correspondence,wang2024amoeba,xiong2024non}. 

A fundamental element of the non-Bloch band theory amounts to replacing the quasimomentum $k$ with $k-i\mu$, or equivalently, extending the lattice momentum to the complex regime. 
As a consequence, the non-Hermitian spectral deformation in the complex-momentum space, by continuously connecting eigenspectra under different boundary conditions (open, periodic, or semi-infinite), establishes a crucial correspondence between complex momenta and non-Hermitian bands,
which has recently led to the Amoeba formulation of the non-Bloch band theory in higher dimensions~\cite{wang2024amoeba}. Such a correspondence 
also provides an intriguing connection between spectral and band topologies in non-Hermitian systems.
Nevertheless, despite the theoretical importance of non-Hermitian spectral deformation, its experimental observation proves to be difficult. 
Previous studies using momentum-resolved spectroscopy~\cite{wang2021generating,patil2022measuring,wang2021topological,tang2020exceptional,zhang2023observation,rao2024braiding}
are insufficient to reveal the non-Bloch features, whereas detection schemes with complex-momentum resolution remain challenging.




In this work, we report the experimental observation of the non-Hermitian spectral deformation in the complex-momentum space. 
The experiment is facilitated by our scalable and programmable platform based on the synthetic orbital-angular-momentum (OAM) dimension of photons inside a degenerate cavity~\cite{yang2023re, yang2022topological,yang2025observing}. 
We implement a highly controllable non-Hermitian lattice with long-range couplings, where real and imaginary components of the complex momenta are respectively encoded in the phase and amplitude modulations of the OAM modes. 
Importantly, we devise a complex-momentum-resolved spectroscopy by
selectively projecting photons through a spatial light modulator (SLM). 
This enables us to extract information on the exceptional points, the Ronkin function and generalized Brillioun zone, as well as the open-boundary spectra. 
Our experiment demonstrates a practical and versatile route to probing non-Bloch physics, and would stimulate further study of non-Hermitian physics in complex-momentum space.

\begin{figure*}
\includegraphics[width=1.6\columnwidth]{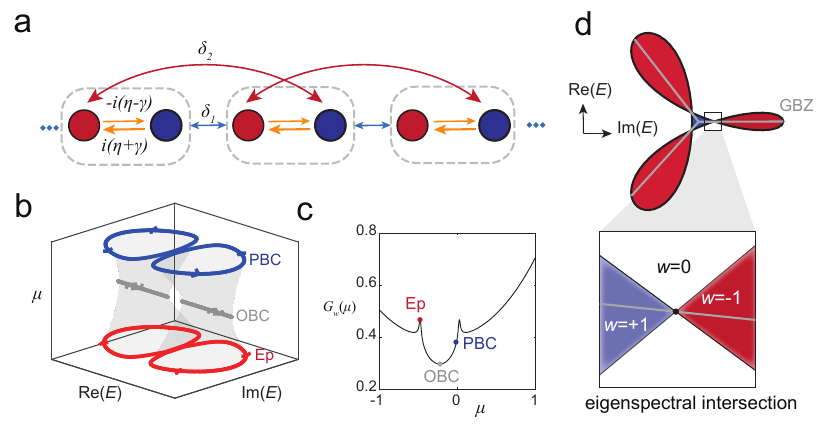}
	\caption{\label{concept} 
\textbf{Spectral geometry and deformation of a non-Hermitian SSH model.} 
\textbf{a.} Schematic of the non-Hermitian SSH lattice with long-range couplings. The gray dashed box indicates a unit cell, and the red and blue spheres denote the sublattice sites.  
\textbf{b.} Evolution of the complex energy spectrum $E(k - i\mu)$ 
with varying imaginary momentum component $\mu$ for $\delta_2=0$. 
Blue loops represent eigenspectra under the periodic boundary condition (PBC), with $\mu=0$. Gray lines indicate those under the open boundary condition (OBC), with $\mu_{\text{GBZ}}=-0.23$. Red loops mark the emergence of exceptional points (EPs) at $\mu=-0.48$, where the line gap closes. The arrows indicate the spectral winding direction as $k$ increases from $0$ to $2\pi$.  
\textbf{c.} The Wasserstein metric $G_w(\mu) = \partial_\mu A(\mu) / 2\pi$, where $A(\mu)$ is the area enclosed by the eigenspectrum. The colored points mark the parameters for the corresponding spectra in \textbf{b}.  
\textbf{d.} Spectral geometry with long-range couplings. {Self-intersections of the spectrum trace the eigenspectrum under OBC (gray dashed), with the corresponding complex momenta moving along the GBZ.} \textit{Inset (magnified view):} self-intersections, which connect the interiors of loops with winding number $w \neq 0$. }
\end{figure*}

\section{Theoretical frameworks}
We focus on a non-Hermitian extended Su–Schrieffer–Heeger (SSH) model with long-range coupling (Fig. \ref{concept}a), with the Hamiltonian
\begin{equation}\label{H0}
\begin{aligned}
H&=-\sum_{m}\big(\delta_1 a_{R,m-1}^{\dag}a_{L,m}+\delta_2 a_{L,m-1}^{\dag}a_{R,m}+\mathrm{h.c.}\big)\\
&\quad - i\sum_{m}\big[(\eta+\gamma)a_{L,m}^{\dag}a_{R,m}-(\eta-\gamma)a_{R,m}^{\dag}a_{L,m}\big],
\end{aligned}
\end{equation}
where $m\in \mathbb{Z}$ labels the lattice sites, 
$a_{R,m}$ and $a_{L,m}$ are the annihilation operators on 
the sublattice sites $R$ and $L$. 
The parameters $\eta$ and $\delta_1$ set the reciprocal intra- and inter-cell couplings, while $\gamma$ represents the non-reciprocal intracell contribution. Long-range hopping terms characterized by $\delta_2$ are also considered which enrich the band structure.
Hamiltonian (\ref{H0}) has been widely studied theoretically as a paradigmatic model for non-Hermitian skin effects and non-Bloch band theory~\cite{yao2018edge,yokomizo2020topological,yin2018geometrical,li2022topological}.


Under a Fourier transform, Hamiltonian (\ref{H0}) becomes $H=\sum_{\beta}\textbf{a}_{\beta}^{\dag}H(\beta)\textbf{a}_{\beta}$, with $\textbf{a}_{\beta}=\frac{1}{\sqrt{N}}\sum_m\textbf{a}_{m}\beta^m$,
$\textbf{a}_{m}=(a_{R,m}, a_{L,m})^T$, and $H(\beta)=E_1(\beta)\sigma_++E_2(\beta)\sigma_-$. Here $E_1(\beta)=-\delta_1\beta^{-1}-\delta_2\beta+i(\eta-\gamma)$ and $E_2(\beta)=-\delta_1\beta-\delta_2\beta^{-1}-i(\eta+\gamma)$ are the off-diagonal terms, and $\sigma_{\pm}=(\sigma_x\pm i\sigma_y)/2$ are the Pauli ladder operators. Crucially, with a complex momentum $K=k-i\mu$ ($k,\mu \in \mathbb{R}$), 
the non-Bloch mode factor $\beta = e^{-iK}$ replaces $e^{-ik}$ in the conventional Bloch waves, and plays a key role in the non-Bloch band theory. 
With varying $\mu$ the eigenspectrum of the Hamiltonian deforms continuously on the complex plane. Such spectral deformation in the complex-momentum space $K$ contains much information on the non-Bloch features of the system.


For instance, in the case of $\delta_2=0$, the spectral deformation continuously connects the eigenspectra of Eq. (\ref{H0}) under different boundary conditions. 
As illustrated in Fig. \ref{concept}b, the spectrum under the periodic boundary condition (PBC) at $\mu=0$ (blue) changes into that under the open boundary condition (OBC) with a negative $\mu_{\text{GBZ}}$ (gray). Further decreasing $\mu$ would lead to the onset of an exceptional point ($E_1(\beta)=0$ or $E_2(\beta)=0$, red), signaling the closing of the line gap. Theoretically, the spectral deformation can be quantized by the Wasserstein metric $G_w(\mu)=\partial_\mu A(\mu)/2\pi$~\cite{xu2025optimal}, which characterizes 
the spectral difference under different $\mu$, and is proportional to the rate of change of the area $A(\mu)$ enclosed by the eigenspectrum (see Methods for details of the Wasserstein metric). As shown in Fig. \ref{concept}c, the position of the local minimum in $G_w(\mu)$ corresponds to the condition of OBC spectrum. It also determines 
the complex momenta ($K=k-i\mu_{\text{GBZ}}$) for the generalized Brillouin zone (GBZ), which is crucial for calculating the non-Bloch topological invariant. Further, the onset of EPs correspond to singularities in the metric (Fig. \ref{concept}c).


On the other hand, switching on the long-range couplings ($\delta_2\neq 0$) leads to more complicated spectral geometries. Nevertheless, by continuously varying $\mu$, the trajectories of the spectral self-intersections naturally outline the eigenspectrum under the OBC (Fig. \ref{concept}d)~\cite{okuma2020topological,hu2025topological}, 
{provided these intersections should also be the connecting points between the interiors of the spectral loops with nonzero spectral winding numbers $w$ (magnified view)~\cite{wu2022connections} (see Methods for definitions of the winding number)}. 


\begin{figure}
\includegraphics[width=1\columnwidth]{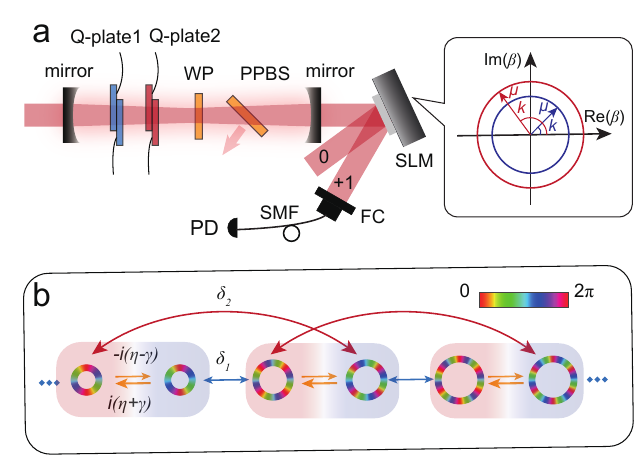}
	\caption{\label{setup} 
\textbf{Experimental implementation and complex-momentum-resolved measurements.} 
\textbf{a.} Schematic of the experimental setup. Two mirrors form a degenerate optical cavity, with the red path indicating the trajectory of the probe light. Colored optical elements inside the cavity introduce controllable mode couplings. A spatial light modulator (SLM) followed by a single-mode fiber (SMF) are used for projective detection. Projection with arbitrary non-Bloch states $\langle\beta^{L}|$ are realized by loading the corresponding phase pattern onto the SLM. \textit{Inset:} 
{For a complete eigenspectra measurement, we scan $\beta$ along circular trajectories with different radius ($\mu$) on the complex plane, and record the output light intensity.}  
WP: wave plate; PPBS: partial polarization beam splitter; FC: fiber coupler; PD: photodetector.  
\textbf{b.} Illustration of the non-Hermitian SSH model encoded in different OAM modes. The colored annuli represent the intensity and phase profiles of the OAM modes. {Red and blue regions denote components with left- and right-circular polarizations, respectively.} The arrows indicate mode couplings, whose colors correspond to the optical elements in \textbf{a} that generate these couplings.
 } 
\end{figure}

\begin{figure*}[t]
\includegraphics[width=2\columnwidth]{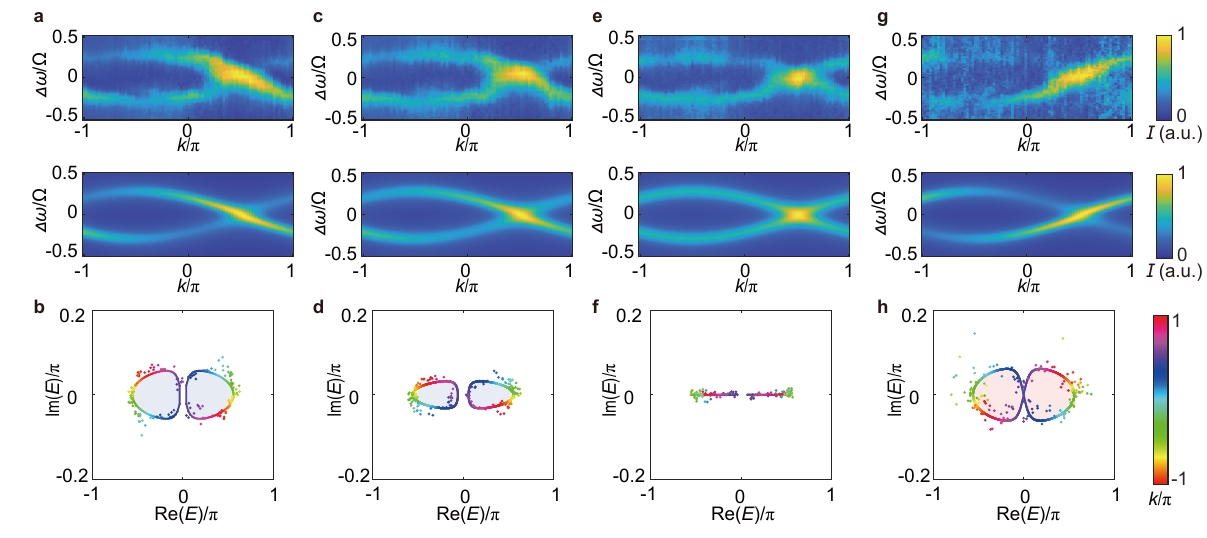}
	\caption{\label{result1} 
\textbf{Measurement of the spectral deformation without long-range couplings.}
\textbf{a, c, e, and g.} Transmission spectra measured at $\mu = 0, -0.1, -0.23, $ and $-0.48$ under the parameters $(\delta_1, \delta_2, \eta, \gamma) = (0.31\pi, 0, 0.25\pi, 0.057\pi)$. The top panel shows the experimental results, and the bottom panel displays the corresponding theoretical predictions. The color scale is normalized to [0, 1]. 
\textbf{b, d, f, and h.} The dots represent the complex eigenenergies extracted from \textbf{a, c, e, and g}, and the solid curves denote the corresponding theoretical calculations. Here $\mu_{\text{GBZ}}=-0.23$ resides on the GBZ, and \textbf{f} corresponds to the eigenspectrum under OBC.
}
\end{figure*}


\section{Experimental realizations}
We employ the synthetic OAM dimension in a degenerate optical cavity to simulate the non-Hermitian SSH model Eq.~\ref{H0}, as illustrated in Fig.~\ref{setup}a. 
The cavity supports a set of resonant OAM modes with vortex phases $e^{-i m \varphi}$ ($m\in \mathbb{Z}$), where each mode serves as a lattice site (Fig.~\ref{setup}b). Here $\varphi$ is the azimuthal angle in the $x$--$y$ plane, with light propagating along $z$. The right- and left-circular polarizations represent the sublattice sites.
More details can be found in Section I of Supplementary Materials (SM).

The intercell coupling is introduced by a Q-plate composed of liquid crystal molecules whose director angle varies azimuthally. Through spin–orbital interaction~\cite{marrucci2006optical}, the Q-plate converts part of a right- (left-) circularly polarized mode into a left- (right-) circularly polarized mode, while changing the OAM index from $m$ to $m+q$. The nearest-neighbor coupling (blue arrows in Fig.~\ref{setup}b) is realized using a Q-plate (with parameter $q=2$, blue elements in Fig.~\ref{setup}a) whose coupling strength $\delta_1$ is electrically tunable. Long-range couplings (red arrows) are realized by inserting an additional Q-plate ($q=-2$, red elements) with strength $\delta_2$. 

The intracell couplings are jointly controlled by a wave plate (WP) and a partial polarizing beam splitter (PPBS). The WP introduces a Hermitian coupling with tunable birefringence delay $\eta$. 
In contrast, the PPBS introduces a skew-Hermitian term by partially reflecting vertically polarized photons out of the cavity, with the reflection rate defining the strength $\gamma$. Together, these two elements (yellow components) generate nonreciprocal intracell couplings (yellow arrows). 
See SM. II for details on the Hamiltonian engieering. 

To directly probe the complex eigenenergies in the synthetic OAM space, we excite the cavity using a tunable continuous-wave (CW) laser. The laser detuning $\Delta\omega$ is scanned relative to the cavity resonance $\Omega$, and the transmitted intensity is synchronously recorded by a photon detector (PD). The detuning $\Delta\omega$ is interpreted as the Floquet quasi-energy of the system. Due to the resonant nature of the cavity, only the eigenmodes of the Hamiltonian are transmitted at resonance, and each transmission peak corresponds to an eigenenergy. The output field contains contributions from all eigenstates, encompassing all momenta.

To resolve the complex momentum-dependent bands, we project the output field onto momentum bases. In the synthetic OAM dimension, the Bloch momentum is conjugate to the lattice site index $m$, corresponding to the azimuthal angle $\varphi$. The Bloch momentum-resolved measurement is achieved by selecting photons with specific azimuthal angles after transmission. However, in the non-Bloch regime, the momentum becomes complex and no longer corresponds to a directly measurable physical quantity, posing an experimental challenge. 

Here we devise a complex-momentum-resolved spectroscopy, using an SLM in combination with a single-mode fiber (SMF). We employ a unique encoding strategy, where a phase-only SLM is used to realize simultaneous amplitude and phase modulation of the optical field (see Methods for details). The SLM converts the non-Bloch component of the transmitted light $E_0= \frac{1}{\sqrt{N}} \sum_{m=1}^{N} \beta^{-m} e^{-i m \varphi}$ into a Gaussian mode in the $+1$ diffraction order, enabling efficient coupling into the SMF for detection. Owing to the biorthogonality of non-Bloch states (see SM. III for details), all other mode components are effectively filtered out. The measurement thus 
projects the output state $|\Psi_{\text{out}}\rangle$ onto a left non-Bloch state, $\langle \beta^{L}|= \frac{1}{\sqrt{N}} \sum_{m=1}^{N} \beta^{-m} \langle m|$, 
{enabling the detection of $|\langle\beta^{L}|\Psi_{\text{out}}\rangle|^2$ which is proportional to the output intensity (see Methods).}
The capability to realize arbitrary complex-valued projective bases is essential for our experiment, and 
and highlights a distinctive advantage of the synthetic OAM dimension.

\begin{figure*}[t]
\includegraphics[width=2\columnwidth]{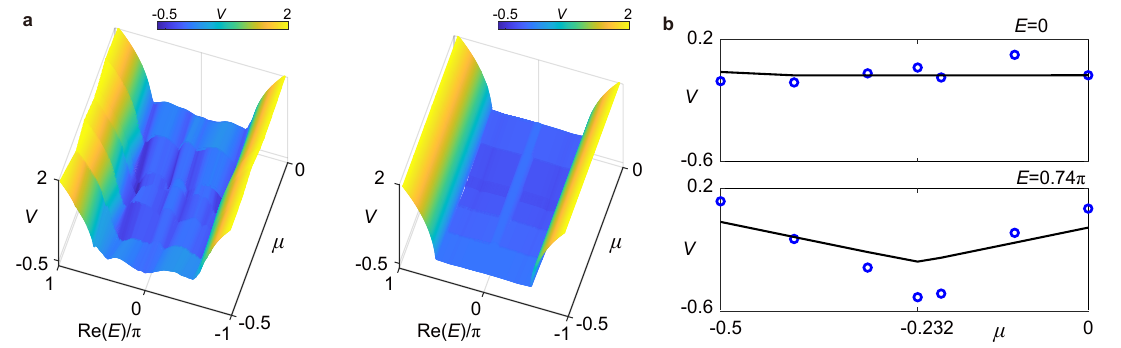}
\caption{\label{resultp} 
\textbf{Ronkin function measurement.}  
\textbf{a} Ronkin-function landscape from experimental data (left) and theoretical calculations (right) under the parameters $(\delta_1, \delta_2, \eta, \gamma) = (0.31\pi, 0, 0.25\pi, 0.057\pi)$.  
\textbf{b} Ronkin function with fixed $E = 0$ and $E = 0.74\pi$, respectively. Experimental data are shown as dots, and theoretical curves are shown as solid lines.
}
\end{figure*}

\begin{figure*}[t]
\includegraphics[width=2\columnwidth]{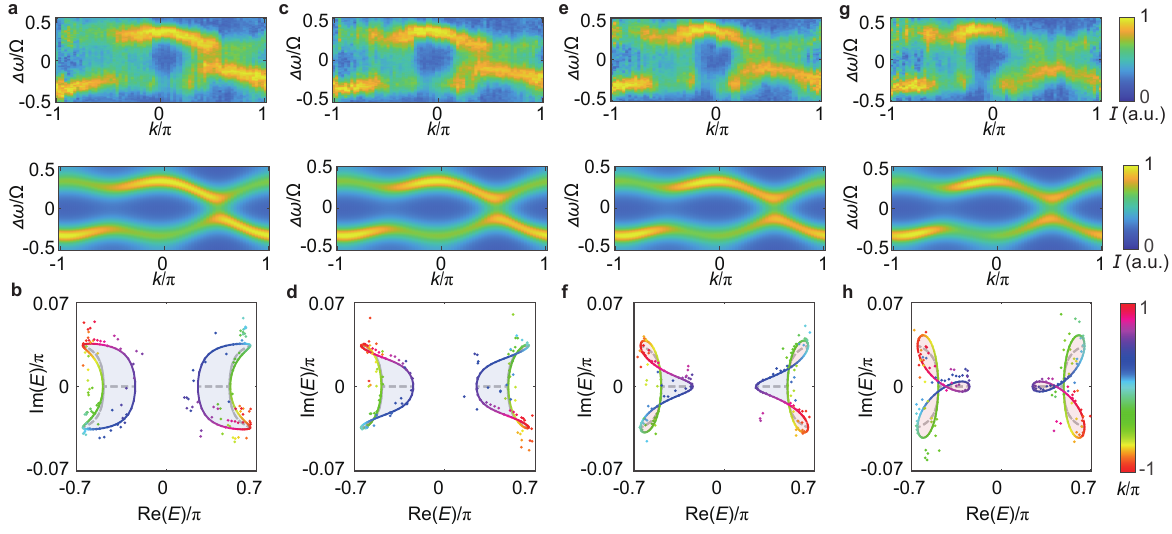}
	\caption{\label{result2} 
\textbf{Measurement of the spectral deformation with long-range couplings.}
\textbf{a, c, e, and g.} Experimental (top) and theoretical (bottom) transmission spectra at $\mu = 0, -0.03, -0.06,$ and $-0.09$, respectively, with $(\delta_1, \delta_2, \eta, \gamma) = (0.13\pi, 0.5\pi, -0.125\pi, 0.036\pi)$.
\textbf{b, d, f, and h.} Dots represent the complex eigenenergies extracted from panels \textbf{a, c, e, and g}; solid curves show the corresponding theoretical results. The thick gray curves denote the eigenspectra under OBC.
} 
\end{figure*}



\section{Spectral deformation.}
To experimentally observe the spectral deformation in Fig.~\ref{concept}, we develop a programmable phase-loading scheme. For a given imaginary component $\mu$ of the complex momentum $K=k-i\mu$, a series of phase masks corresponding $k \in [0, 2\pi)$ are sequentially generated and loaded onto the SLM. The associated non-Bloch mode factor $\beta$ traces a circular contour in the complex plane, as illustrated in Fig.~\ref{setup}a. The eigenenergies under these basis states satisfy biorthogonality and form closed loops in the complex energy plane. By varying $\mu$—that is, changing the radius of the contour—we load another set of phase masks to probe the variation of the spectra in the complex plane. 


Theoretically, the complex-momentum–resolved transmission spectrum is given by
$I(\beta, \Delta\omega) \propto |G(\beta, \Delta\omega)|^2$,
where 
$G(\beta, \Delta\omega) = \sum_s\left[1 - t e^{i(2\pi \Delta\omega/\Omega - E_s(\beta))} \right]^{-1}$ (see Methods for details). Here, $t$ is the field retention coefficient per round trip in the absence of modulation, $s=\pm$ is the band index, and $E_s(\beta) = \text{Re}[E_s(\beta)]+i\text{Im}[E_s(\beta)]$ denotes the complex eigenspectrum of the corresponding band. 
For a given $(k,\mu)$, $\text{Re}[E_s(\beta)]$ and $\text{Im}[E_s(\beta)]$ 
correspond to the detuning and linewidth of the transmission peaks, respectively, and are extracted from the measured transmission spectra.  


We first measure the transmission spectra at $\mu = 0$, corresponding to the Bloch limit under PBC. The coupling parameters are set to $\delta_1 = 0.31\pi$, $\delta_2 = 0$, $\eta = 0.25\pi$, and $\gamma = 0.057\pi$. The measured transmission spectra for different $k$ values are shown in the top panel of Fig.~\ref{result1}a. The bottom panel presents the numerically simulated transmission spectra $I(\beta, \Delta\omega)$, which agree well with the experimental results. The extracted complex eigenenergies are shown as dots in Fig.~\ref{result1}b, over the theoretical bands (solid lines). The two bands form distinct loops in the complex plane, enclosing areas with spectral winding number $w=1$ (light-blue shaded). 

Figs.~\ref{result1}c and~\ref{result1}e show the experimental and theoretical transmission spectra for $\mu=-0.1$ and $\mu_{\text{GBZ}}=-0.23$, respectively. The extracted complex eigenenergies are plotted in Figs.~\ref{result1}d and~\ref{result1}f, which clearly show the spectral deformation from the PBC case (with $\mu=0$) toward the OBC case (at $\mu_{\text{GBZ}}$). In particular, the spectrum becomes purely real under OBC,
wherein the parameter $\mu_{\text{GBZ}}$ manifests itself as a local minimum of the Wasserstein metric, and the GBZ can be constructed from the corresponding set of complex momenta through $\beta=e^{ik+\mu_{\text{GBZ}}}$.


Further decreasing $\mu$ to $-0.48$ ($\mu<\mu_{\rm GBZ}$), the evolution of the transmission spectra is shown in Fig.~\ref{result1}g, and the extracted complex energies are plotted in Fig.~\ref{result1}h. Here the point gap reopens, and the eigenenergy bands enclose regions with a winding number $w = -1$ (light-red shaded). 
Interesting, at $\mu=-0.48$, the line gap closes and an EP emerges, consistent with theoretical prediction. 
We note that another EP appears at $\mu=0.02$, symmetric with respect to $\mu_{\mathrm{GBZ}}$, and the merging of these EPs (under appropriate parameter change) would signal a non-Bloch topological phase transition.

A complementary way to characterize the correspondence between the complex momentum and non-Bloch bands is through the Ronkin function~\cite{wang2024amoeba}, defined as
\begin{align}
V(E,\mu) = \frac{1}{2\pi}\int_0^{2\pi} \ln|\det[E - H(\beta)]|dk.
\end{align}
Using the experimentally measured spectra $E_{\mathrm{exp}}(\beta)$ with $k \in (0, 2\pi]$ and $\mu$ varying from $-0.5$ to $0$, we construct the Ronkin function as
$V(E,\mu) = \frac{1}{N_k} \sum_k \ln|E - E_{\mathrm{exp}}(\beta)|$,
where $N_k$ is the number of sampling points for $k$, and the reference energy $E$ is chosen within $(-\pi, \pi]$. 

In Fig.~\ref{resultp}a, we show the experimentally constructed (left panel) and the numerically calculated (right panel) Ronkin-function landscape as functions of $\text{Re}{(E)}$ and $\mu$. 
Information of the GBZ and OBC spectrum can be obtained from the local minimum of the landscape. As more clearly illustrated in Fig.~\ref{resultp}b for two representative reference energies, when the reference energy $E$ does not belong to the OBC spectrum, the Ronkin function becomes flat as it reaches its minimum (top panel). By contrast, the Ronkin function features a sharp local minimum at $\mu_{\text{GBZ}}$ when the reference energy lies within the OBC spectrum (see also Fig.~\ref{result1}f). Our observations are consistent with prediction of the non-Bloch band theory.


                
Finally, we switch on the long-range couplings $\delta_2$, under which the non-Hermitian bands are enriched by complicated geometries.
At $\mu = 0$, the experimental and theoretical transmission spectra are shown in Fig.~\ref{result2}a, and the corresponding complex energy bands are plotted in Fig.~\ref{result2}b. The OBC spectrum reduces to tree-like arcs (gray dashed lines) lying within the interiors of the two loops (light-blue shaded) with spectral winding $w = 1$.
When $\mu$ decreases, spectral self-intersections start to emerge, as shown in Figs .~\ref{result2}d and f, which are extracted from the transmission spectra Figs .~\ref {result2}c and e. These self-intersections partition the spectrum of each band into three regions: two small loops (light-red shaded) traversed clockwise with winding number $w=-1$, and the large loop (light-blue shaded) traversed counterclockwise with $w=1$. As $\mu$ varies, the self-intersections move along separate branches of the OBC spectrum (gray dashed). When $\mu$ decreases further to $-0.09$ (Figs.~\ref{result2}g and h), a third self-intersection point appears, on the remaining branch of the OBC spectrum. 
Importantly, since the intersection-associated $\beta$ all lie on the GBZ, the GBZ can be reconstructed by tracking the evolution of the spectral self-intersections in the complex-momentum space.



\section{Discussion}
We experimentally observe the non-Hermitian spectral deformation in the complex momentum space of a non-Hermitian SSH model. 
By establishing the correspondence between complex momenta and non-Hermitian bands, our work provides direct experimental support for the non-Bloch band theory, and offers a unique access to non-Bloch features, such as the OBC spectrum, GBZ, EPs and phase transitions, through spectral deformation.  



Our platform exhibits two distinctive advantages. First, the resonant nature of the cavity provides an intrinsic computational advantage. Unlike time-bin~\cite{weidemann2022topological} or path-encoded synthetic lattices~\cite{zhang2025nonchiral}, where eigeninformation is extracted via Fourier transform and state tomography, our system retrieves eigenenergies and non-Bloch features directly from the steady-state transmission. This drastically reduces the computational and storage cost, particularly as system size increases. 
Second, in contrast to the standard momentum-resolved spectroscopies such as angle-resolved photoemission or time-resolved spectroscopy in the synthetic frequency dimension~\cite{dutt2019experimental},
our newly developed scheme of complex-momentum-resolved spectroscopy 
provides access to the complex momentum space, paving the way for future experimental studies of the non-Bloch band theory. Owing to its programmability and scalability, our photonic platform can be readily extended to explore a broad class of non-Bloch physics, including auxiliary generalized Brillouin zones~\cite{yang2020non,wu2022connections}, non-Bloch parity-time symmetry breaking~\cite{hu2024geometric}, and Amoeba formulation~\cite{wang2024amoeba} in higher dimensions. 

When finalizing this work, we noticed a relevant prepint arXiv:2510.20160, where non-Hermitian band structures in the complex momentum space are mapped using acoustic crystals.


\section{Methods}
\textbf{Wasserstein Metric and the Spectral Winding Number}
As $\mu$ varies, the eigenspectrum deforms continuously. To measure the spectra differences and extract non-Bloch structures through spectral geometry, Ref.~\cite{xu2025optimal} introduces the Wasserstein metric $G_w(\mu)$, which is defined as the optimal transport distance between spectra $H(e^{ik+\mu})$ and $H(e^{ik+\mu+\mathrm{d}\mu})$. Assuming that $\{E_i(\beta)\}$ is the set of energy bands of $H(\beta)$, the Wasserstein metric is
\begin{equation}
    G_w(\mu) = \sum_i\int_0^{2\pi} \frac{\mathrm{d}k}{2\pi} \left|\frac{\partial E_i}{\partial k}\right|^2:=\partial_\mu \frac{A(\mu)}{2\pi}.
\end{equation}
Here $A(\mu)$ is the area enclosed by the eigenspectrum.

On the other hand, the spectral winding number $w(E,\mu)$ is defined with respect to a reference energy $E$ for an imaginary flux $\mu$
\begin{equation}
    w(E,\mu) = \int_0^{2\pi}\frac{\mathrm{d}k}{2\pi} \partial_k \log\det [H( \beta)-E].
\end{equation}
Protected by the point gap of the eigenspectrum, the winding number speaks to the topological origin of the non-Hermitian skin effect.

~\\

\textbf{Non-Bloch projective measurements with a phase-only hologram}
The central operation in our experiment is the projection onto a complex momentum state, in the form of the left non-Bloch states $\langle \beta^L| = \frac{1}{\sqrt{N}}\sum_{m=1}^{N} \beta^{-m} \langle m|$.
To realize this projection, we need to design a hologram to convert the target optical field
\begin{equation}
E_o(x,y) = \frac{1}{\sqrt{N}}\sum_{m=1}^{N} \beta^{-m} e^{-im\varphi(x,y)}
\end{equation}
into a plane wave that can be coupled into an SMF. Here the azimuthal angle is defined as $\varphi(x,y) = \tan^{-1}(y/x)$, with $x$ and $y$ denoting the pixel coordinates. Coupling into the SMF effectively implements a projective measurement, as all other spatial modes are filtered out.
 
Since the SLM provides only phase modulation, we need to encode both amplitude and phase information of $E_o(x,y)$ into a phase-only hologram $T_\beta(x,y)$. Inspired by the Ref.~\cite{bolduc2013exact}, here we choose the hologram in the form
\begin{equation}
T_\beta(x,y) = e^{i M(x,y) \, \text{mod}\big[F(x,y) + 2\pi m/\Lambda, 2\pi\big]},
\end{equation}
where $\Lambda$ is the period of a blazed grating that diffracts light into the first order, and $M(x,y)$ and $F(x,y)$ are undetermined modulation functions independent of $E_o(x,y)$. After passing through the SLM, the near-field light becomes $E_0 \cdot T(x,y)$. Using the Fourier expansion, the first-order diffracted field can be written as
\begin{equation}
\begin{aligned}
\Psi_1(x,y)&=-|E_0(x,y)| \, \mathrm{sinc}[\pi M(x,y)] \times \\
&e^{ i \big[ F(x,y) + \pi M(x,y) + \mathrm{Arg}[E_0(x,y)] \big]}.
\end{aligned}
\end{equation}

To ensure efficient coupling into the SMF, the first-order field should be a plane wave, that is, $|\Psi_1(x,y)| = C$ and $\mathrm{Arg}[\Psi_1(x,y)] = 0$, where $C$ is a constant. This condition determines the modulation functions
\begin{equation}
\begin{aligned}
M(x,y) &= 1 + \frac{1}{\pi} \mathrm{sinc}^{-1} \frac{C}{|E_o(x,y)|}, \\
F(x,y) &= -\mathrm{Arg}[E_o(x,y)] - \pi M(x,y).
\end{aligned}
\end{equation}

~\\
\textbf{Input-Output Relations of the Cavity System}
We excite the synthetic OAM lattice by injecting a light field into the cavity, with the input given by
\begin{equation}
|\Psi_{\text{in}}\rangle = |\psi^{\text{in}}_s\rangle \otimes |O_{\text{in}}\rangle.
\end{equation}
Here $|\psi^{\text{in}}_s\rangle$ and $|O_{\text{in}}\rangle$ represent the initial polarization and OAM states, respectively.  
After multiple round trips inside the cavity, the transmitted output field can be expressed as a coherent superposition of all transmitted components
\begin{equation}
|\Psi_{\text{out}}\rangle = -|\kappa|^2 \sum_{l} t^l e^{i 2\pi l \Delta\omega / \Omega} U^l |\Psi_{\text{in}}\rangle,
\end{equation}
where $l$ denotes the round-trip number, $\kappa$ is the coupling coefficient of the mirror, $t$ is the field retention coefficient per round trip, $\Delta\omega$ is the detuning between the input laser and the cavity resonance frequency $\Omega$, and $U = e^{-iH}$ is the single-round-trip evolution operator governed by the effective Hamiltonian $H$. We insert the completeness relation over both polarization and OAM degrees of freedom as
\begin{equation}
\sum_{s,\beta} |\psi^R_s\rangle\langle \psi^L_s| \otimes |\beta^R\rangle\langle \beta^L| = I,
\end{equation}
where $|\psi^R_s\rangle$ and $\langle \psi^L_s|$ denote the right and left eigenstates of $H(\beta)$, respectively. The non-Bloch basis states are defined as 
$\langle \beta^{L}| = \frac{1}{\sqrt{N}} \sum_{m=1}^{N} \beta^{-m} \langle m|$ and 
$|\beta^{R}\rangle = \frac{1}{\sqrt{N}} \sum_{m=1}^{N} \beta^{m} |m\rangle$.  
Substituting this relation into the expression of the output state yields
\begin{equation}
\begin{aligned}
|\Psi_{\text{out}}\rangle
= & -|\kappa|^2 \sum_{l} t^l e^{i 2\pi l \Delta\omega / \Omega} U^l \times \\
&\sum_{s,\beta} |\psi^R_s\rangle\langle \psi^L_s| \otimes |\beta^R\rangle\langle \beta^L| \Psi_{\text{in}}\rangle \\
= & -|\kappa|^2 \sum_{s,\beta} \sum_{l} t^l e^{i 2\pi l \Delta\omega / \Omega} e^{-i l E_s(\beta)} \times\\
&\langle \psi^L_s | \psi^{\text{in}}_s \rangle
\langle \beta^L | O_{\text{in}} \rangle 
|\psi^R_s\rangle |\beta^R\rangle.
\end{aligned}
\end{equation}

Summing over $l$ gives
\begin{equation}
\begin{aligned}
|\Psi_{\text{out}}\rangle = &
-|\kappa|^2 \sum_{s,\beta} 
\frac{1}{1 - t e^{i[2\pi \Delta\omega / \Omega - E_s(\beta)]}} \times\\
&\langle \psi^L_s | \psi^{\text{in}}_s \rangle
\langle \beta^L | O_{\text{in}} \rangle 
|\psi^R_s\rangle |\beta^R\rangle.
\end{aligned}
\end{equation}

As discussed in the previous section, passing through PBS and coupling the output field into the SMF leads to a non-Bloch projective measurement, with the transimitted amplitude
\begin{equation}
\begin{aligned}
\langle \psi^{\text{out}}_s, \beta_{\text{SLM}} | \Psi_{\text{out}}\rangle
&= -|\kappa|^2 \sum_{s,\beta} 
\frac{1}{1 - t e^{i[2\pi \Delta\omega / \Omega - E_s(\beta)]}}\times \\
&\langle \psi^L_s | \psi^{\text{in}}_s \rangle
\langle \psi^{\text{out}}_s | \psi^R_s \rangle
\langle \beta_{SLM}^R |\beta^R\rangle
\langle \beta^L | O_{\text{in}} \rangle.
\end{aligned}
\end{equation}

In the experiment, both the input and output polarization states are chosen as right-circularly polarized, so that
$\langle \psi^L_s | \psi^{\text{in}}_s \rangle 
\langle \psi^{\text{out}}_s | \psi^R_s \rangle = 1/2$.  
The input OAM mode is a Gaussian beam ($m=0$), yielding $\langle \beta^L | O_{\text{in}} \rangle = 1/\sqrt{N}$.  
For the biorthogonality condition of the non-Bloch states, the projected optical field becomes
\begin{equation}
\langle \psi^{\text{out}}_s, \beta_{\text{SLM}} | \Psi_{\text{out}}\rangle
= -\frac{|\kappa|^2}{2\sqrt{N}}
\sum_s 
\frac{1}{1 - t e^{i[2\pi \Delta\omega / \Omega - E_s(\beta_{\text{SLM}})]}}.
\end{equation}

We define the Green's function as
\begin{equation}
G(\beta, \Delta\omega) = 
\sum_s \frac{1}{1 - t e^{i[2\pi \Delta\omega / \Omega - E_s(\beta)]}},
\end{equation}
so that the transmitted intensity can be expressed as
\begin{equation}
I(\beta, \Delta\omega) 
= |\langle \psi^{\text{out}}_s, \beta | \Psi_{\text{out}}\rangle|^2 
\propto |G(\beta, \Delta\omega)|^2.
\end{equation}

~\\
\noindent\textbf{Acknowledgement}\\
This work was supported by the Quantum Science and Technology-National Science and Technology Major Project (Grant Nos.\,2021ZD0301400, and 2021ZD0301200), the National Natural Science Foundation of China (Grants No. 11874343, 92365205, 11974334, 12374479, 12404576, and W2411001), the Postdoctoral Innovative Talents Support program (BX20230349), China Postdoctoral Science Foundation Funded Project (2024M763125), Xiaomi Young Talents Program, the Fundamental Research Funds for the Central Universities (Grant No. WK5290000003), the USTC Major Frontier Research Program (Grant No. LS2030000002).  The authors acknowledge the technical support from the USTC Center for Micro and Nanoscale Research and Fabrication.\\ 

\noindent\textbf{Author contributions}\\
M. Y. and W. Y. conceived the study. 
Y. L. and M.Y. performed the experiments and
processed experimental data. 
M.-T. X. and W. Y. contributed to the theoretical analysis.
J.-S. X., C. -F. L. and G. -C. G. supervised the project.
All authors read the paper and discussed the results.\\

\noindent\textbf{Conflict of interest}\\ 
The authors declare no competing financial interests.\\

\noindent\textbf{Data availability:} All data needed to evaluate the conclusions in the paper are present in the paper and/or the Supplementary Materials.

\normalem
\bibliographystyle{naturemag}
\bibliography{reference}

@article{bansil2016colloquium,
  title={Colloquium: Topological band theory},
  author={Bansil, Arun and Lin, Hsin and Das, Tanmoy},
  journal={Reviews of Modern Physics},
  volume={88},
  number={2},
  pages={021004},
  year={2016},
  publisher={APS}
}

@article{haldane1988model,
  title={Model for a quantum Hall effect without Landau levels: Condensed-matter realization of the" parity anomaly"},
  author={Haldane, F Duncan M},
  journal={Physical review letters},
  volume={61},
  number={18},
  pages={2015},
  year={1988},
  publisher={APS}
}

@article{armitage2018weyl,
  title={Weyl and Dirac semimetals in three-dimensional solids},
  author={Armitage, N Peter and Mele, Eugene J and Vishwanath, Ashvin},
  journal={Reviews of Modern Physics},
  volume={90},
  number={1},
  pages={015001},
  year={2018},
  publisher={APS}
}

@article{xu2015discovery,
  title={Discovery of a Weyl fermion semimetal and topological Fermi arcs},
  author={Xu, Su-Yang and Belopolski, Ilya and Alidoust, Nasser and Neupane, Madhab and Bian, Guang and Zhang, Chenglong and Sankar, Raman and Chang, Guoqing and Yuan, Zhujun and Lee, Chi-Cheng and others},
  journal={Science},
  volume={349},
  number={6248},
  pages={613--617},
  year={2015},
  publisher={American Association for the Advancement of Science}
}

@article{liu2016quantum,
  title={The quantum anomalous Hall effect: theory and experiment},
  author={Liu, Chao-Xing and Zhang, Shou-Cheng and Qi, Xiao-Liang},
  journal={Annual Review of Condensed Matter Physics},
  volume={7},
  number={1},
  pages={301--321},
  year={2016},
  publisher={Annual Reviews}
}

@article{chang2013experimental,
  title={Experimental observation of the quantum anomalous Hall effect in a magnetic topological insulator},
  author={Chang, Cui-Zu and Zhang, Jinsong and Feng, Xiao and Shen, Jie and Zhang, Zuocheng and Guo, Minghua and Li, Kang and Ou, Yunbo and Wei, Pang and Wang, Li-Li and others},
  journal={Science},
  volume={340},
  number={6129},
  pages={167--170},
  year={2013},
  publisher={American Association for the Advancement of Science}
}

@article{shen2018topological,
  title={Topological band theory for non-Hermitian Hamiltonians},
  author={Shen, Huitao and Zhen, Bo and Fu, Liang},
  journal={Physical review letters},
  volume={120},
  number={14},
  pages={146402},
  year={2018},
  publisher={APS}
}

@article{kawabata2019symmetry,
  title={Symmetry and topology in non-Hermitian physics},
  author={Kawabata, Kohei and Shiozaki, Ken and Ueda, Masahito and Sato, Masatoshi},
  journal={Physical Review X},
  volume={9},
  number={4},
  pages={041015},
  year={2019},
  publisher={APS}
}

@article{gong2018topological,
  title={Topological phases of non-Hermitian systems},
  author={Gong, Zongping and Ashida, Yuto and Kawabata, Kohei and Takasan, Kazuaki and Higashikawa, Sho and Ueda, Masahito},
  journal={Physical Review X},
  volume={8},
  number={3},
  pages={031079},
  year={2018},
  publisher={APS}
}

@article{yao2018edge,
  title={Edge states and topological invariants of non-Hermitian systems},
  author={Yao, Shunyu and Wang, Zhong},
  journal={Physical review letters},
  volume={121},
  number={8},
  pages={086803},
  year={2018},
  publisher={APS}
}

@article{yokomizo2019non,
  title={Non-Bloch band theory of non-Hermitian systems},
  author={Yokomizo, Kazuki and Murakami, Shuichi},
  journal={Physical review letters},
  volume={123},
  number={6},
  pages={066404},
  year={2019},
  publisher={APS}
}

@article{kunst2018biorthogonal,
  title={Biorthogonal bulk-boundary correspondence in non-Hermitian systems},
  author={Kunst, Flore K and Edvardsson, Elisabet and Budich, Jan Carl and Bergholtz, Emil J},
  journal={Physical review letters},
  volume={121},
  number={2},
  pages={026808},
  year={2018},
  publisher={APS}
}

@article{wang2021topological,
  title={Topological complex-energy braiding of non-Hermitian bands},
  author={Wang, Kai and Dutt, Avik and Wojcik, Charles C and Fan, Shanhui},
  journal={Nature},
  volume={598},
  number={7879},
  pages={59--64},
  year={2021},
  publisher={Nature Publishing Group UK London}
}

@article{patil2022measuring,
  title={Measuring the knot of non-Hermitian degeneracies and non-commuting braids},
  author={Patil, Yogesh SS and H{\"o}ller, Judith and Henry, Parker A and Guria, Chitres and Zhang, Yiming and Jiang, Luyao and Kralj, Nenad and Read, Nicholas and Harris, Jack GE},
  journal={Nature},
  volume={607},
  number={7918},
  pages={271--275},
  year={2022},
  publisher={Nature Publishing Group UK London}
}

@article{wang2021generating,
  title={Generating arbitrary topological windings of a non-Hermitian band},
  author={Wang, Kai and Dutt, Avik and Yang, Ki Youl and Wojcik, Casey C and Vu{\v{c}}kovi{\'c}, Jelena and Fan, Shanhui},
  journal={Science},
  volume={371},
  number={6535},
  pages={1240--1245},
  year={2021},
  publisher={American Association for the Advancement of Science}
}

@article{okuma2020topological,
  title={Topological origin of non-Hermitian skin effects},
  author={Okuma, Nobuyuki and Kawabata, Kohei and Shiozaki, Ken and Sato, Masatoshi},
  journal={Physical review letters},
  volume={124},
  number={8},
  pages={086801},
  year={2020},
  publisher={APS}
}

@article{tang2020exceptional,
  title={Exceptional nexus with a hybrid topological invariant},
  author={Tang, Weiyuan and Jiang, Xue and Ding, Kun and Xiao, Yi-Xin and Zhang, Zhao-Qing and Chan, Che Ting and Ma, Guancong},
  journal={Science},
  volume={370},
  number={6520},
  pages={1077--1080},
  year={2020},
  publisher={American Association for the Advancement of Science}
}

@article{zhang2023observation,
  title={Observation of acoustic non-Hermitian Bloch braids and associated topological phase transitions},
  author={Zhang, Qicheng and Li, Yitong and Sun, Huanfa and Liu, Xun and Zhao, Luekai and Feng, Xiling and Fan, Xiying and Qiu, Chunyin},
  journal={Physical Review Letters},
  volume={130},
  number={1},
  pages={017201},
  year={2023},
  publisher={APS}
}

@article{rao2024braiding,
  title={Braiding reflectionless states in non-Hermitian magnonics},
  author={Rao, Zejin and Meng, Changhao and Han, Youcai and Zhu, Liping and Ding, Kun and An, Zhenghua},
  journal={Nature Physics},
  volume={20},
  number={12},
  pages={1904--1911},
  year={2024},
  publisher={Nature Publishing Group UK London}
}

@article{yang2023re,
  title={Realization of exceptional points along a synthetic orbital angular momentum dimension},
  author={Yang, Mu and Zhang, Hao-Qing and Liao, Yu-Wei and Liu, Zheng-Hao and Zhou, Zheng-Wei and Zhou, Xing-Xiang and Xu, Jin-Shi and Han, Yong-Jian and Li, Chuan-Feng and Guo, Guang-Can},
  journal={Science Advances},
  volume={9},
  number={4},
  pages={eabp8943},
  year={2023},
  publisher={American Association for the Advancement of Science}
}

@article{yang2022topological,
  title={Topological band structure via twisted photons in a degenerate cavity},
  author={Yang, Mu and Zhang, Hao-Qing and Liao, Yu-Wei and Liu, Zheng-Hao and Zhou, Zheng-Wei and Zhou, Xing-Xiang and Xu, Jin-Shi and Han, Yong-Jian and Li, Chuan-Feng and Guo, Guang-Can},
  journal={Nature Communications},
  volume={13},
  number={1},
  pages={2040},
  year={2022},
  publisher={Nature Publishing Group UK London}
}

@article{yang2025observing,
  title={Observing half-integer topological winding numbers in non-Hermitian synthetic lattices},
  author={Yang, Mu and Liao, Yu-Wei and Zhang, Hao-Qing and Li, Yue and Hao, Zhi-He and Zhou, Zheng-Wei and Luo, Xi-Wang and Xu, Jin-Shi and Li, Chuan-Feng and Guo, Guang-Can},
  journal={Light: Science \& Applications},
  volume={14},
  number={1},
  pages={225},
  year={2025},
  publisher={Nature Publishing Group UK London}
}

@article{yokomizo2020topological,
  title={Topological semimetal phase with exceptional points in one-dimensional non-Hermitian systems},
  author={Yokomizo, Kazuki and Murakami, Shuichi},
  journal={Physical Review Research},
  volume={2},
  number={4},
  pages={043045},
  year={2020},
  publisher={APS}
}

@article{yin2018geometrical,
  title={Geometrical meaning of winding number and its characterization of topological phases in one-dimensional chiral non-Hermitian systems},
  author={Yin, Chuanhao and Jiang, Hui and Li, Linhu and L{\"u}, Rong and Chen, Shu},
  journal={Physical Review A},
  volume={97},
  number={5},
  pages={052115},
  year={2018},
  publisher={APS}
}

@article{li2022topological,
  title={Topological properties in non-Hermitian tetratomic Su-Schrieffer-Heeger lattices},
  author={Li, Jia-Rui and Zhang, Lian-Lian and Cui, Wei-Bin and Gong, Wei-Jiang},
  journal={Physical Review Research},
  volume={4},
  number={2},
  pages={023009},
  year={2022},
  publisher={APS}
}

@article{xu2025optimal,
  title={Optimal spectral transport of non-Hermitian systems},
  author={Xu, Mingtao and Gong, Zongping and Yi, Wei},
  journal={Physical Review B},
  volume={111},
  number={21},
  pages={214305},
  year={2025},
  publisher={APS}
}

@article{hu2025topological,
  title={Topological origin of non-Hermitian skin effect in higher dimensions and uniform spectra},
  author={Hu, Haiping},
  journal={Science Bulletin},
  volume={70},
  number={1},
  pages={51--57},
  year={2025},
  publisher={Elsevier}
}

@article{yao2018non,
  title={Non-hermitian chern bands},
  author={Yao, Shunyu and Song, Fei and Wang, Zhong},
  journal={Physical review letters},
  volume={121},
  number={13},
  pages={136802},
  year={2018},
  publisher={APS}
}

@article{xiao2020non,
  title={Non-Hermitian bulk--boundary correspondence in quantum dynamics},
  author={Xiao, Lei and Deng, Tianshu and Wang, Kunkun and Zhu, Gaoyan and Wang, Zhong and Yi, Wei and Xue, Peng},
  journal={Nature Physics},
  volume={16},
  number={7},
  pages={761--766},
  year={2020},
  publisher={Nature Publishing Group UK London}
}

@article{zhang2020correspondence,
  title={Correspondence between winding numbers and skin modes in non-Hermitian systems},
  author={Zhang, Kai and Yang, Zhesen and Fang, Chen},
  journal={Physical Review Letters},
  volume={125},
  number={12},
  pages={126402},
  year={2020},
  publisher={APS}
}

@article{wang2024amoeba,
  title={Amoeba formulation of non-Bloch band theory in arbitrary dimensions},
  author={Wang, Hong-Yi and Song, Fei and Wang, Zhong},
  journal={Physical Review X},
  volume={14},
  number={2},
  pages={021011},
  year={2024},
  publisher={APS}
}

@article{xiong2024non,
  title={Non-Hermitian skin effect in arbitrary dimensions: non-Bloch band theory and classification},
  author={Xiong, Yuncheng and Xing, Ze-Yu and Hu, Haiping},
  journal={arXiv preprint arXiv:2407.01296},
  year={2024}
}

@article{marrucci2006optical,
  title={Optical spin-to-orbital angular momentum conversion in inhomogeneous anisotropic media},
  author={Marrucci, Lorenzo and Manzo, Carlo and Paparo, Domenico},
  journal={Physical review letters},
  volume={96},
  number={16},
  pages={163905},
  year={2006},
  publisher={APS}
}

@article{yang2020non,
  title={Non-Hermitian bulk-boundary correspondence and auxiliary generalized Brillouin zone theory},
  author={Yang, Zhesen and Zhang, Kai and Fang, Chen and Hu, Jiangping},
  journal={Physical Review Letters},
  volume={125},
  number={22},
  pages={226402},
  year={2020},
  publisher={APS}
}

@article{wu2022connections,
  title={Connections between the open-boundary spectrum and the generalized Brillouin zone in non-Hermitian systems},
  author={Wu, Deguang and Xie, Jiao and Zhou, Yao and An, Jin},
  journal={Physical Review B},
  volume={105},
  number={4},
  pages={045422},
  year={2022},
  publisher={APS}
}

@article{hu2024geometric,
  title={Geometric origin of non-Bloch PT symmetry breaking},
  author={Hu, Yu-Min and Wang, Hong-Yi and Wang, Zhong and Song, Fei},
  journal={Physical Review Letters},
  volume={132},
  number={5},
  pages={050402},
  year={2024},
  publisher={APS}
}

@article{dutt2019experimental,
  title={Experimental band structure spectroscopy along a synthetic dimension},
  author={Dutt, Avik and Minkov, Momchil and Lin, Qian and Yuan, Luqi and Miller, David AB and Fan, Shanhui},
  journal={Nature communications},
  volume={10},
  number={1},
  pages={3122},
  year={2019},
  publisher={Nature Publishing Group UK London}
}

@article{zhang2025nonchiral,
  title={Nonchiral Non-Bloch Invariants and Topological Phase Diagram in Nonunitary Quantum Dynamics without Chiral Symmetry},
  author={Zhang, Yue and Li, Shuai and Xu, Yingchao and Tian, Rui and Zhang, Miao and Li, Hongrong and Gao, Hong and Zubairy, M Suhail and Li, Fuli and Liu, Bo},
  journal={Physical Review Letters},
  volume={134},
  number={11},
  pages={113603},
  year={2025},
  publisher={APS}
}

@article{weidemann2022topological,
  title={Topological triple phase transition in non-Hermitian Floquet quasicrystals},
  author={Weidemann, Sebastian and Kremer, Mark and Longhi, Stefano and Szameit, Alexander},
  journal={Nature},
  volume={601},
  number={7893},
  pages={354--359},
  year={2022},
  publisher={Nature Publishing Group UK London}
}

@article{bolduc2013exact,
  title={Exact solution to simultaneous intensity and phase encryption with a single phase-only hologram},
  author={Bolduc, Eliot and Bent, Nicolas and Santamato, Enrico and Karimi, Ebrahim and Boyd, Robert W},
  journal={Optics letters},
  volume={38},
  number={18},
  pages={3546--3549},
  year={2013},
  publisher={Optical Society of America}
}

\end{document}